1 **Divergence in age-patterns of mortality change drives**

2 **international divergence in lifespan inequality**

5 Duncan O. S. Gillespie[1,2]*, Meredith V. Trotter[1] and Shripad D. Tuljapurkar[1]

7 [1]*Department of Biology, Stanford University, Stanford, CA 94305, USA*

8 [2]*Department of Public Health & Policy, University of Liverpool, Liverpool, L69 3GB,*

9 *UK*

10 *Corresponding author:

11 Email. d.gillespie@liverpool.ac.uk

12 Tel: +44 (0)151 795 0436

13 Fax: +44 (0)151 794 5588

**In the past six decades, lifespan inequality has varied greatly within and among countries even while life expectancy has continued to increase. How and why does mortality change generate this diversity? We derive a precise link between changes in age-specific mortality and lifespan inequality, measured as the variance of age at death. Key to this relationship is a young–old threshold age, below and above which mortality decline respectively decreases and increases lifespan inequality. First, we show that shifts in the threshold's location modified the correlation between changes in life expectancy and lifespan inequality over the last two centuries. Second, we analyze the post Second World War trajectories of lifespan inequality in a set of developed countries, Japan, Canada and the United States (US), where thresholds centered on retirement age. Our method reveals how divergence in the age-pattern of mortality change drives international divergence in lifespan inequality. Most strikingly, early in the 1980s, mortality increases in young US males led lifespan inequality to remain high in the US, while in Canada the decline of inequality continued. In general, our wider international comparisons show that mortality change varied most at young working ages after the Second World War, particularly for males. We conclude that if mortality continues to stagnate at young ages, yet declines steadily at old ages, increases in lifespan inequality will become a common feature of future demographic change.**

# Introduction

Lifespan inequality is the defining measure of social and health disparity and, alongside life expectancy (the mean age at death), is a key indicator of population health (Tuljapurkar et al. 2000; Edwards and Tuljapurkar 2005). Changes in life expectancy are well understood in terms of the underlying changes in mortality: mortality reduction at any age increases life expectancy (Keyfitz 1977; Goldman and Lord 1986; Vaupel 1986; Vaupel and Canudas-Romo 2003). However, the precise link between lifespan inequality and age-specific mortality is less clear. First, only mortality decline at ages below a young–old threshold can decrease lifespan inequality (Zhang and Vaupel 2009; Vaupel et al. 2011; van Raalte and Caswell 2013). Second, there are several ways to quantify lifespan inequality and understand its change through time (Wilmoth and Horiuchi 1999; Cheung et al. 2005; Edwards and Tuljapurkar 2005; Horiuchi et al. 2008; Vaupel et al. 2011). Although highly correlated, these measures behave differently in response to change in the age structure of mortality (van Raalte and Caswell 2013). Here, we measure lifespan inequality by the variance of age at death, which measures the dispersion in age at death relative to life expectancy (Edwards and Tuljapurkar 2005). The variance of age at death is a central parameter in population and economic modeling (Tuljapurkar 2008; Caswell 2009; Tuljapurkar and Edwards 2011; Edwards 2012; Schindler et al. 2012) and it is therefore important to understand how it responds to mortality change. We present an exact relationship between age-specific mortality and the variance of age at death (our Eq. 2 below), which is a more transparent and demographically interpretable relationship than that derived by Caswell (Eq. 42 in Caswell (2009)). Our framework provides a powerful tool for quantifying how trajectories of age-specific mortality affect the variance, both within countries and in an international sample of countries.

Given period age-specific mortality, $\mu(x)$, and the corresponding probability of survival to each age $l(x)$, the age-distribution of death is $\varphi(x) = \mu(x)l(x)$. Figure 1 contrasts the age-distribution of death in Sweden in 1800 and 2000, illustrating the standard outcomes of development: (i) reduction of child death; (ii) increase of life expectancy; (iii) emergence and advance of a modal age at adult death; (iv) decrease of lifespan inequality (Lee and Carter 1992; Bongaarts 2005; Canudas-Romo 2008; Vaupel et al. 2011). These arise from an initial focus of mortality decline on children and young adults followed by a shift of mortality decline to older ages (Cutler et al. 2006).

Since about 1950 in the developed countries, life expectancy has increased consistently but change in lifespan inequality among countries has varied greatly (Edwards and Tuljapurkar 2005; Peltzman 2009; Smits and Monden 2009; Engelman et al. 2010). This is in striking contrast to the strong negative correlation between changes in life expectancy and lifespan inequality present during the first century of mortality decline (Vaupel et al. 2011). The recent variability of this relationship, we show, results from the very different responses of life expectancy and lifespan inequality to age-specific mortality change. First, we show that in Sweden, the threshold for the variance of age at death increased from childhood in the late 18$^{th}$ century to around 65 years (y) in 1950. This movement of the threshold altered the relationship between changes in life expectancy and lifespan inequality. Second, we examine the pattern of post Second

World War lifespan inequality across many countries, and then focus on Japan, Canada and the United States (US). The application of our decomposition to these post war trajectories of lifespan inequality showed that fluctuations in lifespan inequality resulted primarily from mortality change at young working ages.

## Results

**The analysis of change**

The variance of age at death after a specified index age $A$ is

$$V(A) = \frac{1}{l(A)} \int_A^\infty \varphi(x)[x - M(A)]^2 dx, \qquad (1)$$

where $M(A)$ is the mean age at death after age $A$ and is equal to the life expectancy after age $A$ plus $A$, i.e. $M(A) = e(A) + A$. If we set $A=0$, then $M(A) = e(0)$, the life expectancy at birth and the variance, $V(0)$, quantifies the dispersion among deaths at all ages. However, for developed countries it can be more informative to focus on adult ages, e.g. using $V(15)$[1].

What is the change in $V(A)$ due to a proportional mortality reduction at an age $x$ greater than $A$? Starting from Eq. 1, we find

---

[1] $V(0)$ is necessarily dominated by the (now usually small) fraction of infant deaths. If we set $A=15$, the mean $M(15)$ and the variance $V(15)$ describe the dispersion of adult deaths. The advantage of focusing on adult ages is that the effects of infant mortality change are removed. Infant mortality has fallen steadily, driving steady declines in $V(0)$ (Edwards and Tuljapurkar 2005; Tuljapurkar and Edwards 2011). Considering only mortality after particular index ages $A$ can often reveal quite different trajectories of $V(A)$ (Engelman et al. 2010). Even with low infant mortality, the inclusion of the few youngest ages can obscure the lifespan inequality effects of adult mortality change. This makes $V(15)$ an ideal measure for investigating how the variability in age-specific mortality trajectories contributes to lifespan inequality change.

$$\frac{dV(A)}{d\ln\mu(x)} = \frac{-2\mu(x)}{l(A)} \int_x^\infty l(z)[z - M(A)]dz, \tag{2}$$

(for details, see Supplementary Information). The integral in Eq. 2 has a negative term: the contribution from ages between $x$ and $M(A)$, and a positive term: the contribution from ages above $M(A)$. If mortality decreases at an age above $M(A)$, the variance $V(A)$ will increase. Now suppose that we reduce mortality at an age below $M(A)$. It is clear from Eq. 2 that reducing mortality at younger ages will gradually cause the negative contribution to dominate, thus decreasing $V(A)$. We therefore define a young–old threshold age, which we call $T(A)$, before which mortality decline decreases $V(A)$, and after which mortality decline increases $V(A)$. The threshold age $T(A)$ is always less than $M(A)$ because the accumulated increases in $V(A)$ incurred above $M(A)$ must be fully offset by decreases below $M(A)$ before reduction in $V(A)$ can occur.

Hereafter, we refer to the rate of change in Eq. 2 as the sensitivity of $V(A)$. Figure 2a shows the sensitivity of the variance for $A=15$, for Sweden in 1950 and 2011. Three features of the plot are typical of mortality in developed countries: (i) mortality decline at ages from 15 to about 65 y decreases $V$; (ii) mortality decline after about 65 y increases $V$; (iii) there is a threshold age, $T(15)$, at approximately 65 y, which separates 'young' ages, where mortality decline decreases $V$, from 'old' ages, where mortality decline increases $V$. The sensitivity of life expectancy $e(15)$ to a proportional mortality decline at age $x$ (Goldman and Lord 1986; Vaupel 1986; Vaupel and Canudas-Romo 2003) is shown for comparison in Figure 2b. The obvious distinctions are that mortality decline at any age increases life expectancy; and that mortality decline close to the threshold has least effect on $V$. The great difference in the shape of these sensitivities of life expectancy

1    and lifespan inequality, especially below the threshold age, is the critical factor driving

2    their divergent responses to the same age structure of mortality change.

4    **Long-term mortality change**

5    Sweden was the first country to record ages at death systematically, with records from

6    1751 to the present (Human Mortality Database 2012). These data provide the best

7    opportunity to understand the long-term association of the young–old threshold with

8    changes in lifespan inequality. Figure 3 displays the annual change in the thresholds $T(A)$,

9    for index ages $A$=0, 15, and 65 y. Figure 4 shows the concurrent changes in $e(A)$ and

10   $V(A)$.

12   To provide some basic intuition on how the threshold depends on the age structure of

13   mortality, we note that a lower bound on the threshold for $A$=0 (derived from Eq. 2 in the

14   Supplementary Information) is

15   $T(0) \geq 2e(0) - \omega$,                    (3)

16   where $\omega$ is the longest lifespan. In the early years of transition to low mortality, before

17   the late 19th century, the lower bound defined by Eq. 3 was below or close to zero. In the

18   first half of the 20th century, life expectancy increased rapidly, which substantially

19   increased the threshold age—e.g. with $e(0)$=65 and $\omega$=100 we have a lower bound at 30

20   y. More recently, life expectancy has begun to converge on $\omega$. This convergence is

21   commonly known as 'rectangularization' of the age structure (Wilmoth and Horiuchi

22   1999), and it causes the lower bound on the threshold to advance. We now turn to exact

23   values of the young–old threshold.

Before about 1875, the threshold $T(0)$ followed a fluctuating increase to about 20 y (Fig. 3), so only mortality decline in children and very young adults could decrease $V$. However, $V(0)$ changed little (Fig. 4a), indicating an even distribution of mortality decline below and above the threshold age. From 1875 to about 1950, the threshold advanced to near 60 y, so mortality decline at a greater range of young ages could now decrease $V$. This reinforced the strong negative correlation between changes to life expectancy and lifespan inequality. In this period, $V$ decreased rapidly and life expectancy increase accelerated, indicating much faster mortality declines below the threshold age. After 1950, the threshold continued advancing to 70–75 y. In this recent period, decreases in $V$ slowed, despite continued increases in life expectancy, indicating the switch to more even mortality decline below and above the threshold age.

Increasing the index age after which we measure lifespan inequality advances the young-old threshold. To investigate changes in lifespan inequality among adults, we use an index of 15 y. $V(15)$ did not begin to decrease rapidly until ca. 1925 (Fig. 4b), when the threshold $T(15)$ was 57 y (Fig. 3). That this decrease began around 50 y later than the decrease in $V(0)$ indicates the much later spread of rapid mortality decline from below 15 y to 15–57 y. By 1950, mortality reduction below 15 y had converged the thresholds $T(0)$ and $T(15)$ to 60–65 y. This initiated a period when the distribution of mortality change between working and retired ages was critical in shaping the trajectory of lifespan inequality.

Consider now an index age of 65 y: the obvious feature of mortality change is the consistent positive relationship of changes in life expectancy and lifespan inequality (Fig. 4c). The threshold $T(65)$ was always near 70 y, so only mortality decline at a few young ages could decrease $V(65)$. Therefore, even with mortality reduction by an equal proportion at each age, there was a strong bias for variance increase. The only deviation occurred after 1980, when the trajectory of $V(65)$ flattened while $e(65)$ increased; the threshold was 75 y in 1980. Comparison of mortality change at ages 65–75 y, and >75 y showed an abrupt increase in mortality decline among males, which was greatest at 65–75 y (Fig. S1). This was large enough to overwhelm the bias for variance increase and, for the first time since 1751, disrupted the positive correlation between increases in life expectancy and lifespan inequality.

**The young–old threshold for mortality in developed countries**

Now we focus on mortality in developed countries after the Second World War, when deaths became concentrated at older adult ages. This gave the age-distribution of death its characteristic 'modern' pattern, with a well-defined modal age in adulthood (Fig. 1). Before investigating the age-specific drivers of lifespan inequality, we present an approximation to the threshold age that is specific to modern mortality. The modern mortality age-pattern is a reasonable fit to the Gompertz model in which mortality increases with age as

$$\mu(x) = Be^{kx}, \tag{4}$$

1   where *B* is the initial level of mortality and *k* is the exponential rate of mortality increase.

2   From Eq. 4 (see Supplementary Information and Tuljapurkar and Edwards (2011)) the

3   mode (*m*) and standard deviation (*σ*) of age at death are

4   $$m \approx \frac{1}{k} \ln \frac{k}{B},$$  (5)

5   and

6   $$\sigma \approx \frac{1}{k}.$$  (6)

7   Life expectancy is given to a good approximation by

8   $$e(0) \approx m - \frac{\sigma}{2},$$  (7)

9   and our threshold is approximated by

10  $$T(0) \approx m - \frac{3\sigma}{2}.$$  (8)

11  This places the threshold at 1.5 standard deviations of the age at death below the modal

12  age, and (using Eq. 7) 1 standard deviation below life expectancy. The threshold's

13  advance is therefore driven by increases in both adult life expectancy and modal age at

14  death.

16  **Age-specific drivers of lifespan inequality change in developed countries**

17  We now focus on the post Second World War changes to *V*(15) in developed countries,

18  for which the young–old threshold was near ages 60–70 y. Our analysis shows that

19  changes in lifespan inequality were shaped by striking differences in mortality change for

20  young versus old adults. We begin with a brief international overview: Figure 5 (solid

21  lines) shows the average trajectories of *V*(15) for each sex across 40 regional data series from the Human Mortality Database (Vaupel et al. 2011). $V(15)$ fluctuated more for males, and the annual directions of $V(15)$ change were also less regionally consistent in males (Fig. S2). Our decomposition (dashed lines) showed that the variance effects of mortality change at young ages (below the threshold) fluctuated more than at old ages (above the threshold), particularly in males. So, fluctuations in $V(15)$ clearly come from mortality change at young ages.

We now examine in more detail Japan, Canada and the United States (US), which illustrate the main features of post-1947 mortality change in developed countries: relatively steady increase in life expectancy (Fig. S3), but large fluctuation in $V$ (Fig. S4). We focus on the slowing of variance decrease and, in some regions, onset of variance increase in the 1980s. The US in particular showed a striking variance increase in the early 1980s. The threshold ages were close to the conventional age of retirement, although there was a steady increase in thresholds over time—by a maximum of 28 y in Japanese females (Fig. S5).

Figure 6 shows how mortality change at young ages (below the threshold) and at old ages (above the threshold) generated the contrasting variance trajectories of each country[2]. In Japan, each sex followed the same basic trajectory of variance change: a rapid decrease from 1947 to about 1975, and increase after about 1990 (Fig. 6a & b). The sexes differed in that $V$ initially decreased faster for females, and the reversal from variance decrease to increase was earlier for females. Our decomposition showed that the initial variance decrease before 1977 was driven by the variance contracting effects of mortality decline at young ages, which dominated the effects of mortality decline at old ages (Fig. 6a & b). Around the 1980s, these opposing effects reached a rough equilibrium and so $V$ changed little. However, after about 1990, the negative effects of mortality decline at young ages weakened further, leading the positive effects of mortality decline at old ages to dominate. This shift in the focus of mortality decline to old ages consequently led $V$ to increase (Fig. S7).

---

[2] We also computed the components of variance change from below and above the fixed age of 65 y (Fig. S6). These components were near identical to those either side of the shifting threshold, indicating that mortality change at ages near the threshold had minimal influence.

*Canada and the US*

In Canada and the US, the variance changes after 1947 were much smaller than in Japan (note the difference in *y*-axis scales in Fig. 6). Prior to 1983, Canada and the US followed similar trajectories, each with little net variance change. However, after 1983, $V$ increased and continued to fluctuate in the US, but began a rapid and consistent decrease in Canada. The result was that, from 1983 to 2007, the Canadian variance (for females and males respectively) change from 8% and 9% lower than the US, to 20% and 15% lower.

We now focus on the period 1983–1994, which captures the major variance divergence between Canada and the US (Fig. 6 c–f). In Canada, mortality decline at young ages produced percentage variance changes of −10.8% for females and −10.1% for males (Fig. 6c & d). From old ages the changes were 1.0% in females and 2.9% in males. In the US, mortality change at young ages produced variance changes of −3.3% in females and 3.6% in males, and from old ages 1.5% and 4.8% respectively (Fig. 6e & f). Thus, due mainly to differences in male mortality change, the US produced a larger component of variance increase from old ages, and a much smaller component of variance decrease (and for males, variance increase) from young ages.

How did mortality change differ at young ages between Canada and the US? In the US, male mortality increased at ages 15–22 and 27–45 y, with increases greater than 30% at 34–37 y (Fig. 7). US females showed a similar pattern, but of smaller magnitude (a maximum increase of 20% at 35 y). By contrast, mortality increase for Canadian males was limited to ages 32–40 y, with a maximum increase of 17%; mortality generally declined for Canadian females. The age-specific changes in mortality therefore differed most between Canada and the US at young working ages, particularly in males. As a result, variance diverged sharply between Canada and the US. When looked at in terms of life expectancy this divergence is almost completely missed (Fig. S3). Thus, diagnosing mortality trends based on life expectancy alone is inadequate and lifespan inequality should feature routinely in demographic monitoring.

## Discussion

We have shown how change in lifespan inequality, as measured by $V$, depends strongly on the balance of mortality change around a young–old threshold. The threshold advances with increases in life expectancy, the modal age at death, and with the compression of mortality into a narrower range of ages. Before about 1950, changes to life expectancy and lifespan inequality had a strong negative correlation (Smits and Monden 2009; Vaupel et al. 2011). This arose because the emphasis of mortality decline was on ages below the threshold (Vaupel et al. 2011). Since 1950, lifespan inequality decrease has slowed, and occasionally even reversed. We show that the post Second World War fluctuations lifespan inequality were driven by fluctuations in mortality at young working ages, below the threshold age, while at old ages mortality decline was relatively consistent. The implication of these trends is clear: if mortality at young ages continues to fluctuate, and mortality at old ages continues its steady decline, lifespan inequality increases will become more common.

But what drives these trajectories of age-specific mortality, causing them to vary both within and among countries? This question brings the focus of our attention to the social determinants of mortality (Graham 2009), and so to the realization that international divergence in lifespan inequality must largely be due to divergence in age-patterns of the social determinants. Mortality is ultimately associated with education, income, social support (from family or public institutions, including the provision of healthcare), lifestyle, disease and living conditions. These interact with the proximate, health related, causes of death at each age to generate lifespan inequality (Edwards and Tuljapurkar 2005; Shkolnikov et al. 2011; Nau and Firebaugh 2012). For example, Japanese development after 1947, including the start of universal healthcare in 1961, generated rapid mortality decline at mainly working ages. It is possible that this rapid progress against mortality gradually left the remaining causes of death in young adults in the category 'hard to prevent'. However, society—and hence the array of social determinants—was also changing, characterized by widening differences in mortality by education and income, particularly in working aged males (Fukuda et al. 2004;

Kagamimori et al. 2009). This could also have slowed mortality declines in young adults, e.g. if progress against the social determinants became limited to the most educated or affluent.

A great deal of research is aimed at understanding why young adult mortality remains high in the US compared to other high-income countries (Crimmins et al. 2011; IOM 2012). One source may be high geographic disparity in young adult mortality within the US. For example, Cullen et al. (2012) showed that mortality differences below age 70 among US counties and racial groups resulted largely from differences in education and income, particularly for males (see also Backlund et al. 2007; Crimmins et al. 2009). Current evidence also suggests that life expectancy differences between educational groups in the US have widened in recent decades (Olshansky et al. 2012). A part of this variation is likely due to differences in healthcare access among young adults (Crimmins et al. 2011). Indeed, any explanation of trends in US mortality must address the hypothesis that the correlation of social deprivation with low healthcare access raises the sensitivity of working-age mortality to macro-economic change. In the US, healthcare access at working ages depends largely on employer-provided health insurance; in 1994, 18.6% of non-elderly adults in the US had no insurance (Holahan and Kim 2000). Being without insurance limits care, particularly preventive care at young ages (Lasser et al. 2006), with evidence for consequently higher mortality (Wilper et al. 2009). The 1980s lifespan inequality divergence between Canada and the US could be explained by such social factors: It is not possible to exclude the explanation that macro-economic change resulted in more people in the US in poor health and with no social safety net to ensure appropriate healthcare (Brenner and Mooney 1983; Siddiqi and Hertzman 2007). A slightly different understanding is gained by looking at the common causes of death at young, working, ages (Harper et al. 2007; Nau and Firebaugh 2012). Harper et al. (2007) show that in the 1980s US, mortality in young adults rose and the black-white gap in life expectancy increased. This was due to differential rises in mortality from human immunodeficiency virus (HIV), heart disease, and homicide. HIV became prevalent in the 1980s and early 1990s, particularly among males aged 25–64 y, more so in the US than Canada (Armstrong et al. 1999; UNAIDS 2013). Nau and Firebaugh (2012) also highlight heart disease and homicides as key contributers to the high US variance of age at death. Heart disease is a leading cause of premature death and accidents and homicides, although being a less common cause, account for a high proportion of deaths at the youngest adult ages where, as we show, the variance is highly sensitive to mortality change.

*Alternate inequality metrics*

There are several established metrics of lifespan inequality (van Raalte and Caswell 2013). For the same mortality data, the young–old threshold can differ widely among these metrics, e.g. by around 20 y between $V$ and the metric known as lifespan disparity or $e^{\dagger}$ (Zhang and Vaupel 2009). Our results indicate that the major post Second World War mortality changes occurred outside the zone of threshold disagreement among metrics, indicating that our qualitative conclusions are robust to the choice of metric. We use $V$ for several reasons: it is easily understood and interpreted as a measure of relative dispersion, easily decomposed (Edwards and Tuljapurkar 2005; Edwards 2011; Nau and Firebaugh 2012), and finds direct use in analyzing the demographic and economic consequences of mortality change (Tuljapurkar 2008; Edwards 2012; Schindler et al. 2012).

*The future*

The young–old threshold in developed countries is now approaching 75–80 y (Figure 2a), so only mortality decline at the oldest retired ages can increase lifespan inequality. Given the ongoing fluctuation of social gradients in age-specific health and mortality, and the importance of social gradients for generating lifespan inequality (Cullen et al. 2012; Olshansky et al. 2012), we suggest that future work focus on the population-level effects of mortality change within specific social groups (Edwards 2011; Shkolnikov et al. 2011; van Raalte et al. 2011; Nau and Firebaugh 2012; van Raalte et al. 2012). The great good of development has been the consistent improvement of medical technology and care for the elderly, reflected in steadily declining old-age mortality (Vaupel 2010). However, our results suggest that without comparable progress against young adult mortality, through action on the social determinants, mortality change is unlikely to follow the historical pattern of rising life expectancy and falling lifespan inequality. Hence, we should be prepared for increases in lifespan inequality to become a more common feature of demographic change.

**Materials and Methods**

We downloaded our data from the Human Mortality Database in January 2013 (Human Mortality Database 2012), see the Background Information for each country,

www.mortality.org. For international comparisons, we used the same data series as Vaupel et al. (2011). Table S1 shows the range of years between 1947 and 2011 covered by these data. We conducted all calculations in the R environment (R Development Core Team 2012). The Supplementary Information gives the derivation of each formula presented. For our age-decomposition of change in $V(15)$, we adapted Eq. 2 to

$$\frac{dV(15)}{dt} = -2\int_{15}^{\infty}\frac{a(x)\mu(x)}{l(15)}\int_{x}^{\infty}l(z)[z - M(15)]dzdx, \qquad (9)$$

which gives the change in $V(15)$ over time interval $dt$. The observed proportional changes in age-specific mortality, $a(x)$, between years $t$ and $t+1$ were computed as

$$a(x) = \frac{\mu(x,t+1) - \mu(x,t)}{\mu(x,t)}. \qquad (10)$$

To convert time in these formulae to discrete 1 y intervals, we discretized the probability density function of age at death, considered survivorship to the mid-point of each age-interval, and substituted instantaneous mortality for central death rates. We counted age in 1 y intervals starting from 0.5 y. Where we present mortality change over wider age intervals, we compute a weighted average of the 1 y central death rates in each interval, using the probability of survival to each age as weights (Ahmad et al. 2001).

## Acknowledgements

This project is funded by the National Institutes of Health grants AG22500 and AG039345 to S.T. and by the Stanford Center for Population Research.

# References

Ahmad, O., C. Boschi-Pinto, A.D. Lopez, C.J.L. Murray, R. Lozano, and M. Inoue. 2001. "Age standardization of rates: a new WHO standard." Geneva: World Health Organization.

Armstrong, G.L., L.A. Conn, and R.W. Pinner. 1999. "Trends in infectious disease mortality in the United States during the 20th century." *Jama-Journal of the American Medical Association* 281(1):61-66.

Backlund, E., G. Rowe, J. Lynch, M.C. Wolfson, G.A. Kaplan, and P.D. Sorlie. 2007. "Income inequality and mortality: A multilevel prospective study of 521248 individuals in 50 US states." *International Journal of Epidemiology* 36(3):590-596.

Bongaarts, J. 2005. "Long-range trends in adult mortality: Models and projection methods." *Demography* 42(1):23-49.

Brenner, M.H.and A. Mooney. 1983. "Unemployment and health in the context of economic change." *Social Science & Medicine* 17(16):1125-1138.

Canudas-Romo, V. 2008. "The modal age at death and the shifting mortality hypothesis." *Demographic Research* 19:1179-1204.

Caswell, H. 2009. "Stage, age and individual stochasticity in demography." *Oikos* 118(12):1763-1782.

Cheung, S.L.K., J.M. Robine, E.J.C. Tu, and G. Caselli. 2005. "Three dimensions of the survival curve: Horizontalization, verticalization, and longevity extension." *Demography* 42(2):243-258.

Crimmins, E.M., J.K. Kim, and T.E. Seeman. 2009. "Poverty and biological risk: The earlier "aging" of the poor." *Journals of Gerontology Series a-Biological Sciences and Medical Sciences* 64(2):286-292.

Crimmins, E.M., S.H. Preston, and B. Cohen. 2011. "Explaining divergent levels of longevity in high-income countries." Panel on understanding divergent trends in longevity in high-income countries; National Research Council.

Cullen, M.R., C. Cummins, and V.R. Fuchs. 2012. "Geographic and racial variation in premature mortality in the U.S.: Analyzing the disparities." *Plos One* 7(4):e32930.

Cutler, D., A. Deaton, and A. Lleras-Muney. 2006. "The determinants of mortality." *Journal of Economic Perspectives* 20(3):97-120.

Edwards, R.D. 2011. "Changes in world inequality in length of life: 1970-2000." *Population and Development Review* 37(3):499-528.

Edwards, R.D. 2012. "The cost of uncertain life span." *Journal of Population Economics*:1-38.

Edwards, R.D.and S. Tuljapurkar. 2005. "Inequality in life spans and a new perspective on mortality convergence across industrialized countries." *Population and Development Review* 31(4):645-674.

Engelman, M., V. Canudas-Romo, and E.M. Agree. 2010. "The implications of increased survivorship for mortality variation in aging populations." *Population and Development Review* 36(3):511-539.

Fukuda, Y., K. Nakamura, and T. Takano. 2004. "Municipal socioeconomic status and mortality in Japan: sex and age differences, and trends in 1973-1998." *Social Science & Medicine* 59(12):2435-2445.

Goldman, N. and G. Lord. 1986. "A new look at entropy and the life table." *Demography* 23(2):275-282.

Graham, H. 2009. "Health inequalities, social determinants and public health policy." *Policy and Politics* 37(4):463-479.

Harper, S., J. Lynch, S. Burris, and G. Davey Smith. 2007. "Trends in the black-white life expectancy gap in the United States, 1983-2003 " *Journal of the American Medical Association* 297(11):1224-1232.

Holahan, J. and J. Kim. 2000. "Why does the number of uninsured Americans continue to grow?" *Health Affairs* 19(4):188-196.

Horiuchi, S., J.R. Wilmoth, and S.D. Pletcher. 2008. "A decomposition method based on a model of continuous change." *Demography* 45(4):785-801.

Human Mortality Database. 2012. "University of California, Berkeley (USA), and Max Planck Institute for Demographic Research (Germany). Available at www.mortality.org or www.humanmortality.de."

IOM. 2012. *How far have we come in reducing health disparities?: Progress since 2000: Institute of Medicine workshop summary*. Washington, DC: The National Academies Press.

Kagamimori, S., A. Gaina, and A. Nasermoaddeli. 2009. "Socioeconomic status and health in the Japanese population." *Social Science & Medicine* 68(12):2152-2160.

Keyfitz, N. 1977. *Applied mathematical demography*. New York: A Wiley-Interscience publication.

Lasser, K.E., D.U. Himmelstein, and S. Woolhandler. 2006. "Access to care, health status, and health disparities in the United States and Canada: Results of a cross-national population-based survey." *American Journal of Public Health* 96(7):1300-1307.

Lee, R.D. and L.R. Carter. 1992. "Modeling and forecasting United-States mortality." *Journal of the American Statistical Association* 87(419):659-671.

Nau, C. and G. Firebaugh. 2012. "A new method for determining why length of life is more unequal in some populations than in others." *Demography*:1-24.

Olshansky, S.J., T. Antonucci, L. Berkman, R.H. Binstock, A. Boersch-Supan, J.T. Cacioppo, B.A. Carnes, L.L. Carstensen, L.P. Fried, D.P. Goldman, J. Jackson, M. Kohli, J. Rother, Y. Zheng, and J. Rowe. 2012. "Differences in life expectancy due to race and educational differences are widening, and many may not catch up." *Health Affairs (Project Hope)* 31(8):1803-1813.

Peltzman, S. 2009. "Mortality Inequality." *Journal of Economic Perspectives* 23(4):175-190.

R Development Core Team. 2012. "R: A language and environment for statistical computing." Vienna, Austria: R Foundation for Statistical Computing.

Schindler, S., S. Tuljapurkar, J.-M. Gaillard, and T. Coulson. 2012. "Linking the population growth rate and the age-at-death distribution." *Theoretical Population Biology* 82(4):244-252.

Shkolnikov, V.M., E.M. Andreev, Z. Zhang, J. Oeppen, and J.W. Vaupel. 2011. "Losses of expected lifetime in the United States and other developed countries: Methods and empirical analyses." *Demography* 48(1):211-239.

Siddiqi, A. and C. Hertzman. 2007. "Towards an epidemiological understanding of the effects of long-term institutional changes on population health: A case study of Canada versus the USA." *Social Science & Medicine* 64(3):589-603.

Smits, J. and C. Monden. 2009. "Length of life inequality around the globe." *Social Science & Medicine* 68(6):1114-1123.

Tuljapurkar, S. 2008. "Mortality declines, longevity risk and aging." *Asia-Pacific Journal of Risk and Insurance* 3(1):37-51.

Tuljapurkar, S. and R.D. Edwards. 2011. "Variance in death and its implications for modeling and forecasting mortality." *Demographic Research* 24:497-525.

Tuljapurkar, S., N. Li, and C. Boe. 2000. "A universal pattern of mortality decline in the G7 countries." *Nature* 405(6788):789-792.

UNAIDS. 2013. *http://www.unaids.org/en/dataanalysis/datatools/aidsinfo/*.

van Raalte, A.A. and H. Caswell. 2013. "Perturbation analysis of indices of lifespan variability." *Demography* in press.
23

van Raalte, A.A., A.E. Kunst, P. Deboosere, M. Leinsalu, O. Lundberg, P. Martikainen, B.H. Strand, B. Artnik, B. Wojtyniak, and J.P. Mackenbach. 2011. "More variation in lifespan in lower educated groups: evidence from 10 European countries." *International Journal of Epidemiology* 40(6):1703-1714.

van Raalte, A.A., A.E. Kunst, O. Lundberg, M. Leinsalu, P. Martikainen, B. Artnik, P. Deboosere, I. Stirbu, B. Wojtyniak, and J.P. Mackenbach. 2012. "The contribution of educational inequalities to lifespan variation." *Population Health Metrics* 10.

Vaupel, J.W. 1986. "How change in age-specific mortality affects life expectancy." *Population Studies* 40(1):147-157.

Vaupel, J.W. 2010. "Biodemography of human ageing." *Nature* 464(7288):536-542.

Vaupel, J.W. and V. Canudas-Romo. 2003. "Decomposing change in life expectancy: A bouquet of formulas in honor of Nathan Keyfitz's 90th birthday." *Demography* 40(2):201-216.

Vaupel, J.W., Z. Zhang, and A.A. van Raalte. 2011. "Life expectancy and disparity: an international comparison of life table data." *BMJ open* 1(1):e000128.

Wilmoth, J.R. and S. Horiuchi. 1999. "Rectangularization revisited: Variability of age at death within human populations." *Demography* 36(4):475-495.

Wilper, A.P., S. Woolhandler, K.E. Lasser, D. McCormick, D.H. Bor, and D.U. Himmelstein. 2009. "Health insurance and mortality in US adults." *American Journal of Public Health* 99(12):2289-2295.

Zhang, Z. and J.W. Vaupel. 2009. "The age separating early deaths from late deaths." *Demographic Research* 20:721-729.

1 **Figures**

2 **Fig. 1** The period age-distributions of death for years 1800 and 2000 in Sweden. The

3 vertical lines show life expectancy at birth, $e(0)$. The modal age at death in adulthood

4 became prominent only after the major process of mortality decline

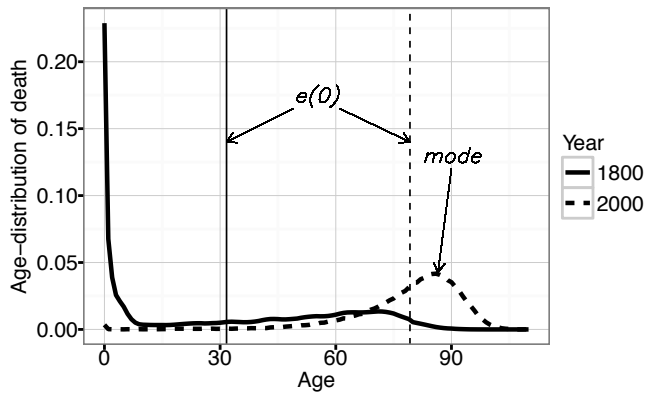

7 **Fig. 2** The age-specific sensitivities to proportional mortality reduction in Sweden for

8 1950 (solid lines) and 2011 (dashed lines) of: (a) the variance of age at death after 15 y;

9 and (b) the mean age at death, i.e. life expectancy, after this age. The vertical lines show

10 the mean age at death after 15 y, $M(15)$, which equals the life expectancy at age 15 y plus

11 15 y, $e(15)+15$

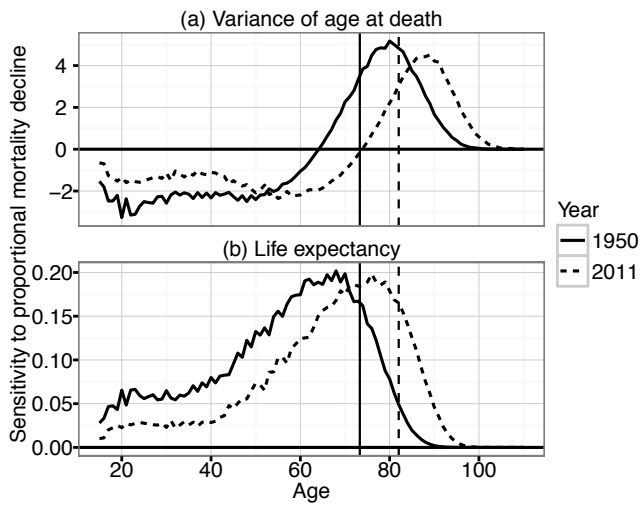

2  **Fig. 3** The young–old thresholds, *T*(*A*) for *A*=0, 15 and 65 y from 1751 to 2011 in

3  Sweden. Note that even in 2011, *T*(65) was only 14 years above age 65 y, a considerably

4  smaller gap than for *T*(0) and *T*(15)

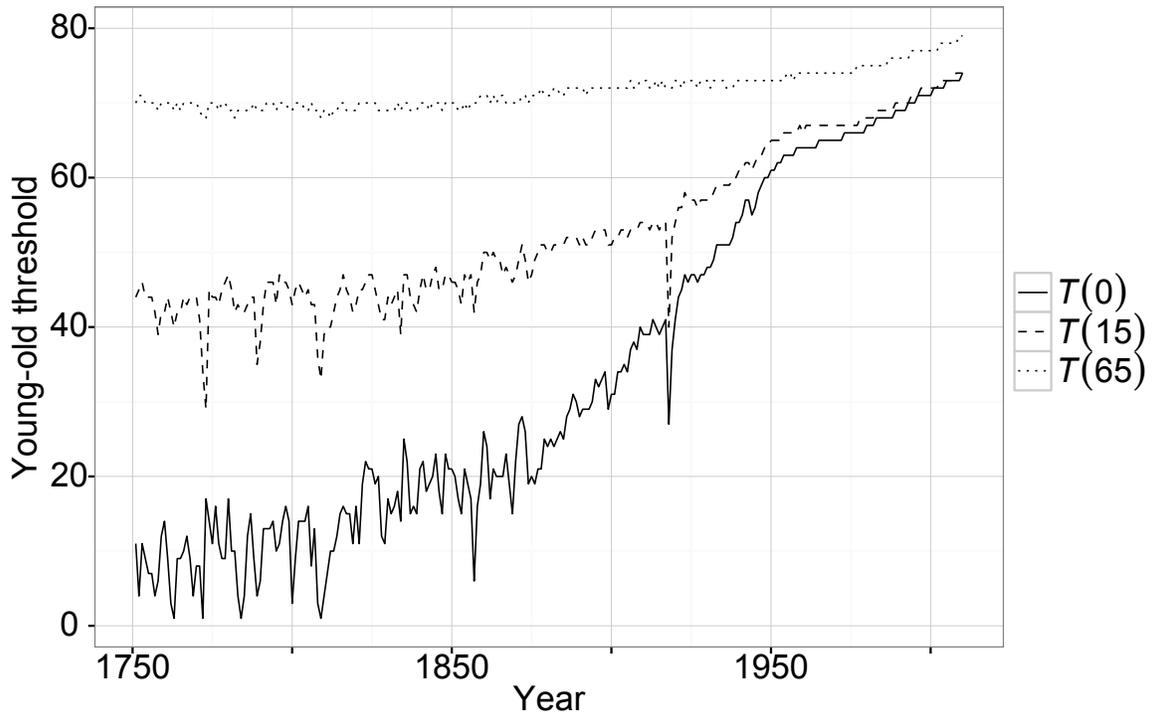

1  **Fig. 4** The trajectories of the variance *V*(*A*) for *A*=0, 15 and 65 y, in panels (a), (b) and (c)

2  respectively, alongside the corresponding trajectories of life expectancy, *e*(*A*). Each

3  trajectory has been scaled so that it ranges between zero and one during the study period

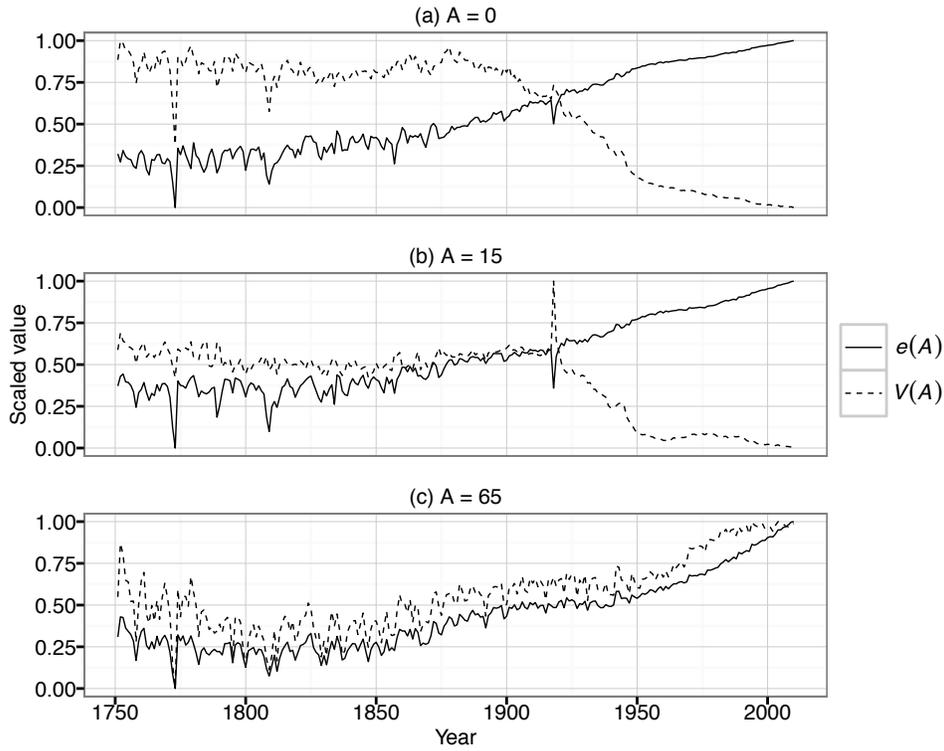

4
5
6
7
8
9
10
11
12
13

1 **Fig. 5** For females (a) and males (b), solid lines show the average trajectory of the

2 variance of age at death after 15 y, $V(15)$, across 40 regions in the Human Mortality

3 Database (see Table S1 for details). The upper dotted lines show this average trajectory as

4 it would have been only with mortality change above the threshold age, $T(15)$. The lower

5 dashed lines show this for mortality change below the threshold. Note that the lines do

6 not show the variance change at each time $t$ ($1947 \le t \le 2011$), but show the cumulative

7 effects of mortality change from 1947 to time $t$. Thus, if $W$ is the annual change in $V$ due

8 to mortality change below or above the threshold in each year, the dotted and dashed

9 lines are given by $V(15, 1947) + \int_{1947}^{t} W(15,u)\,du$. The grey shading around the dotted and

10 dashed lines shows the annual 95% range of variability among regions in each component

11 of change

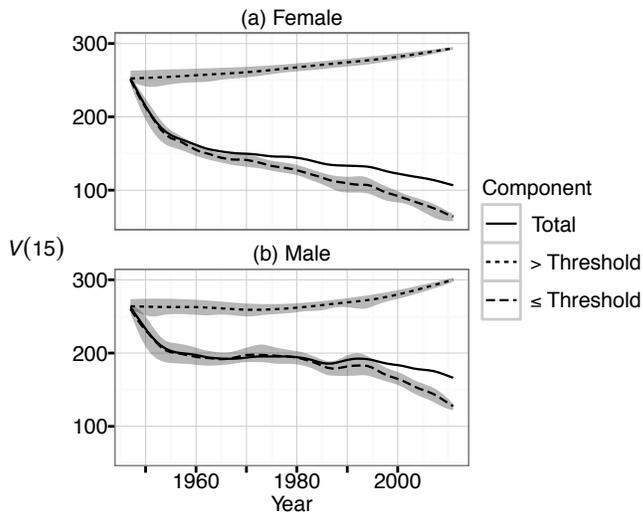

1  **Fig. 6** Decomposition of the variance of age at death after 15 y, $V(15)$ into the effects of
2  mortality change below and above the young–old threshold age, $T(15)$. Solid lines show
3  the overall variance change, dashed lines the change produced from ages below the
4  threshold, and dotted lines the change produced from ages above the threshold. We show
5  separate decompositions for females and males in Japan (a & b), Canada (c & d) and the
6  United States (e & f) from 1947 to 2007. Lines show the cumulative variance effects of
7  mortality change plus $V(15, 1947)$, see details in the legend of Fig. 5

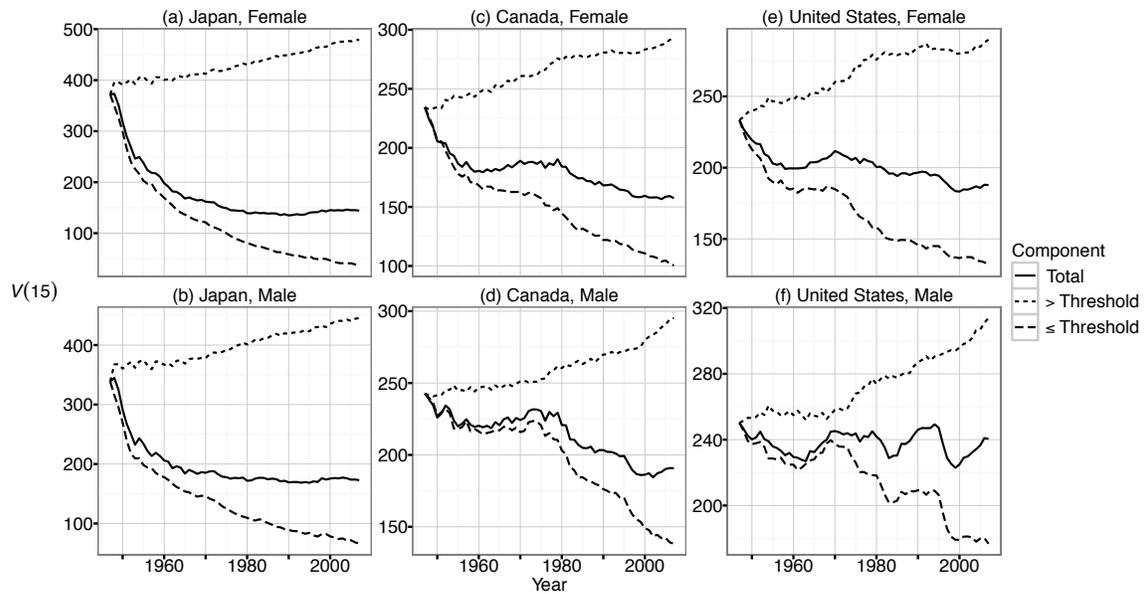

1  **Fig. 7** The percentage changes to the central one-year death rates at ages from 15 years to

2  the average young–old threshold for females (71 years) and males (64 years) between

3  1983 and 1994 in (a) Canada and (b) the United States. Separate lines show the 5-year

4  moving averages of the percentage change in females and males

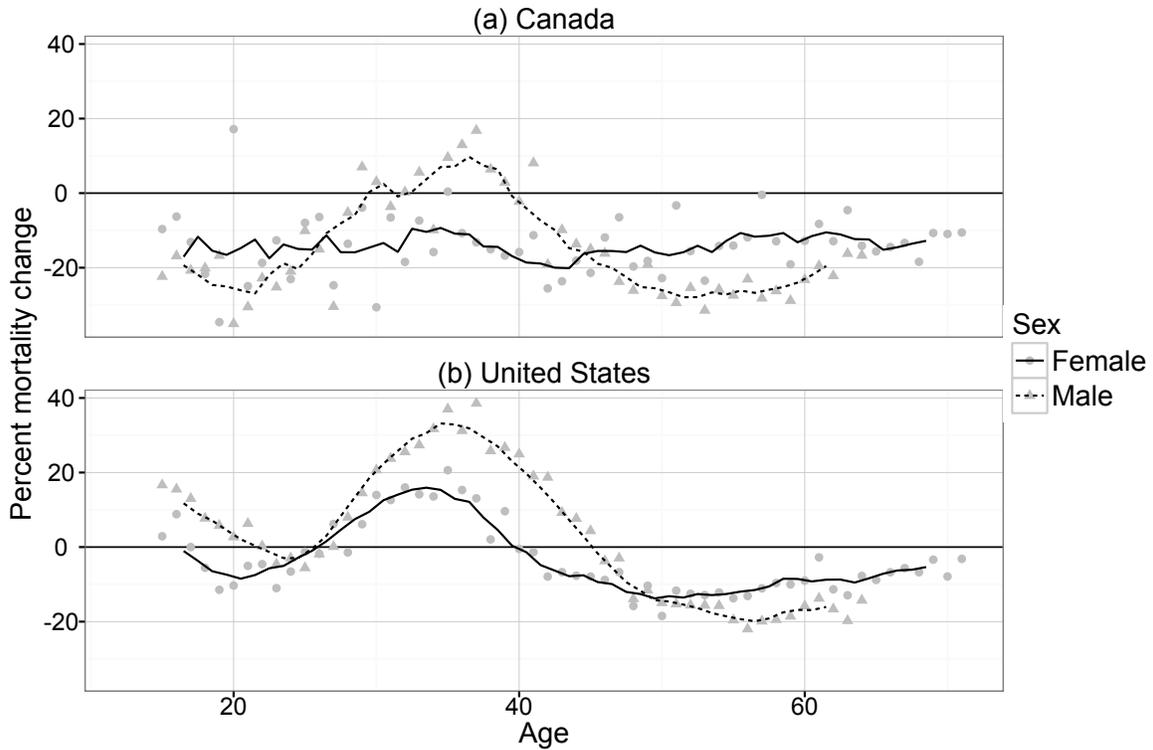

1   **Fig. S1** Weighted average one-year central death rates between 65–75 years and above 75

2   years for each sex in Sweden from 1950 to 2010. The weights used in the calculation

3   were proportional to the probability of survival to each age. The trajectories have been

4   standardized so that each ranges between zero and one

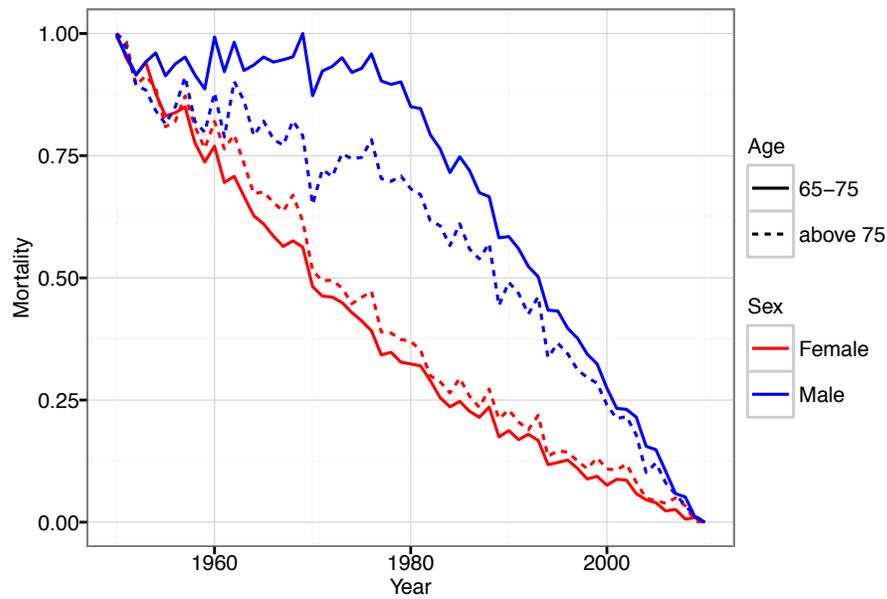

**Fig. S2** The average trajectories of the variance of age at death after 15 years, $V(15)$, in 40 regions from the Human Mortality Database for females (solid lines) and males (dashed lines). The grey shading shows the 95% range among regions of the direction of variance change in each year. The maximum extent of years is 1947–2011, but not all regions have data for each year; see details in Table S1

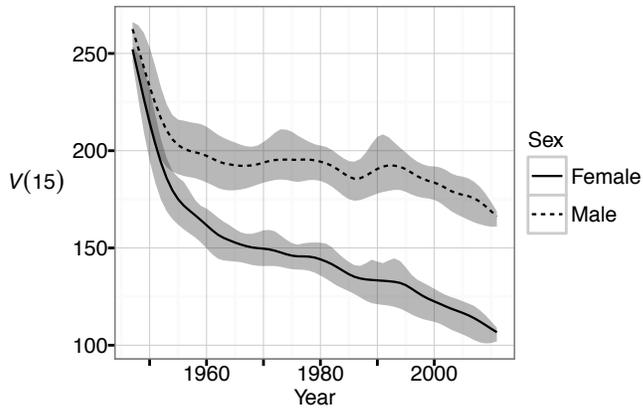

1  **Fig. S3** Life expectancy after zero years i.e. at birth, *e*(0) and 15 years, *e*(15) respectively

2  in (a) & (b) Japan, (c) & (d) Canada and (e) & (f) the United States from 1947 to 2007.

3  Separate lines show the total population, females only and males only

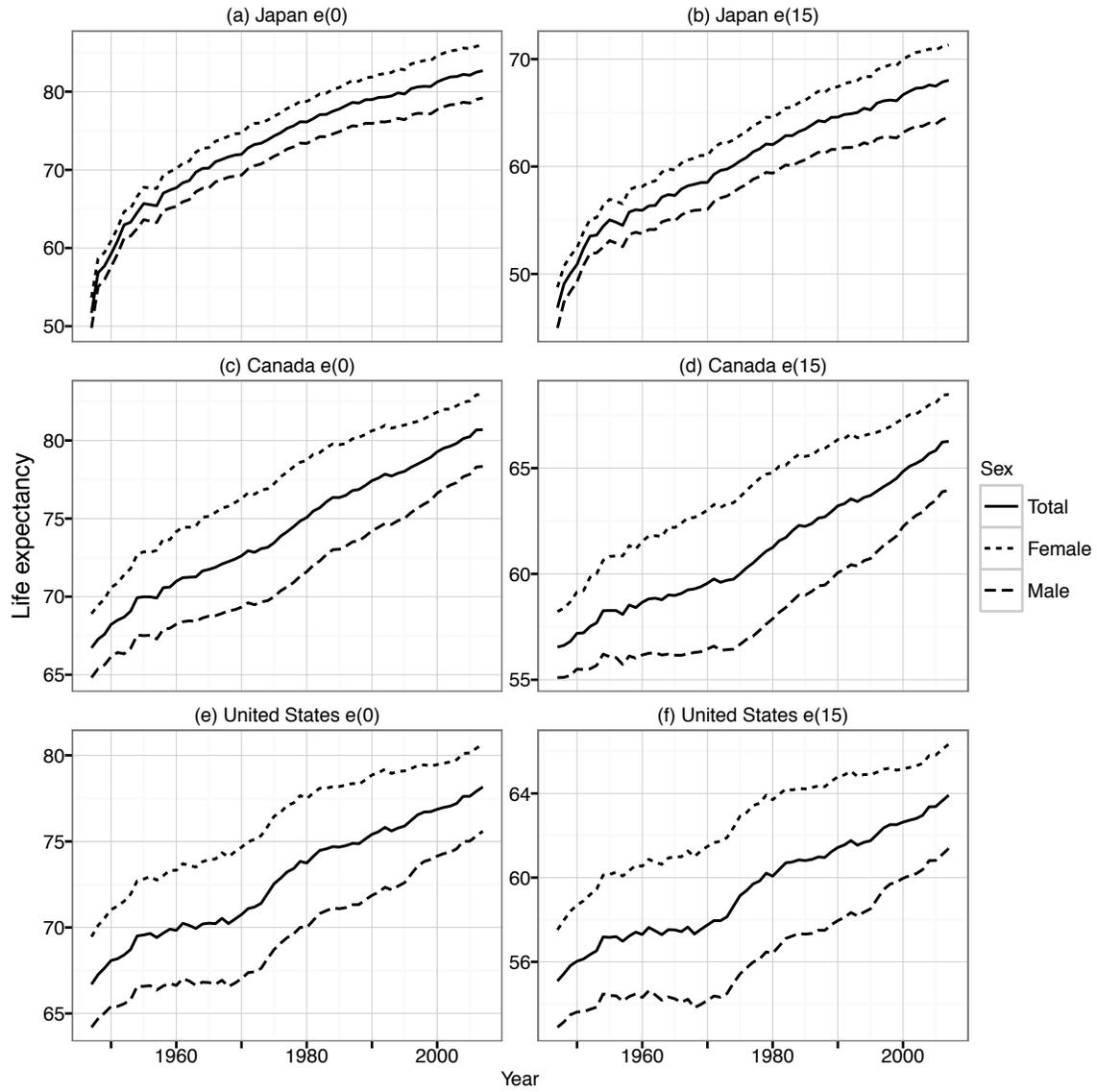

1   **Fig. S4** The variance of age at death after 15 years, $V(15)$ in (a) Japan, (b) Canada and (c)

2   the United States from 1947 to 2007. Separate lines show the total population, females

3   only and males only

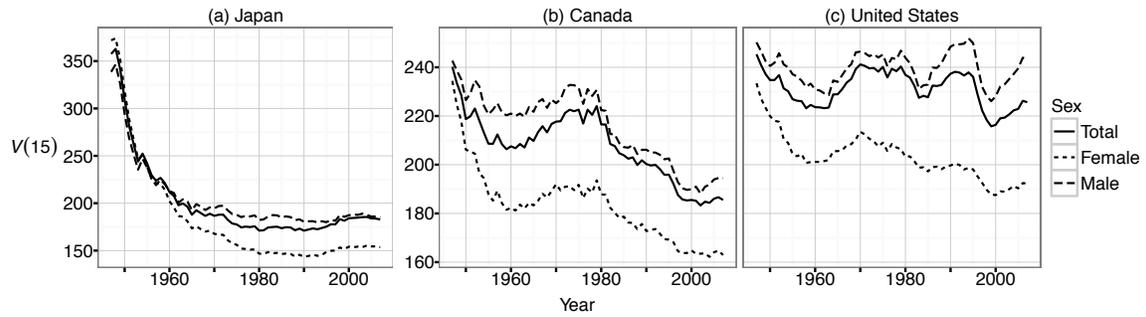

**Fig. S5** The young–old threshold for individuals alive at 15 years, $T(15)$, for (a) Japan, (b) Canada and (c) the United States from 1947 to 2007. Separate lines show the total population, females only and males only. The blue lines show the fit of a loess smooth to the integer values that arise from our use of mortality data in discrete one-year age and period form

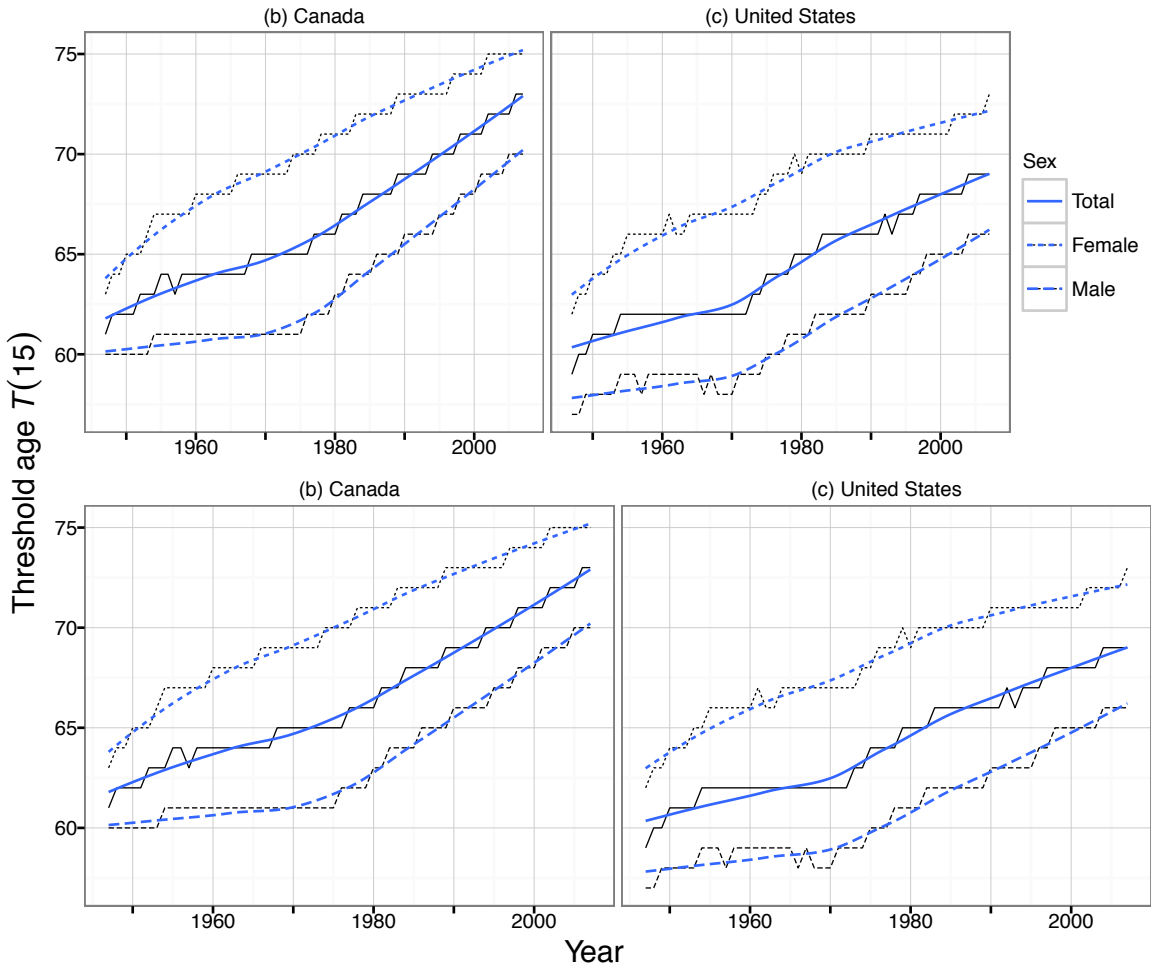

**Fig. S6** Decomposition of the variance of age at death after 15 years, $V(15)$ into the effects of mortality change at working ages (15–65 years) and retired ages (>65 years). Solid lines show the total cumulative change, dashed lines the cumulative change produced by working ages only, and dotted lines the cumulative change produced by retired ages only. We computed separate decompositions for females and males in Japan (a & b), Canada (c & d) and the United States (e & f) from 1947 to 2007

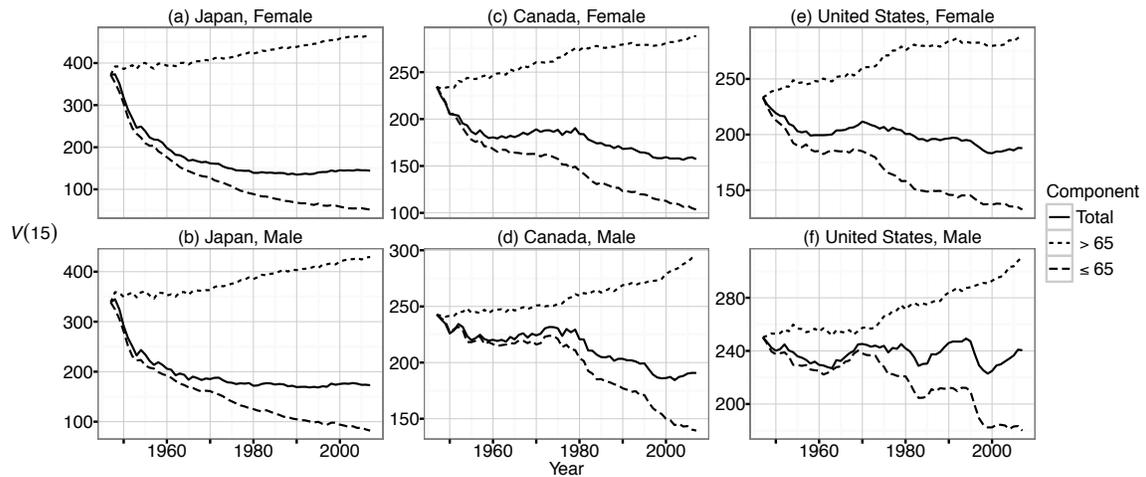

**Fig. S7** Weighted average one-year central death rates between 15–65 years and above 65 years for each sex in Japan from 1947 to 2007. The weights used in the calculation were proportional to the probability of survival to each age. The trajectories have been standardized so that each ranges between zero and one

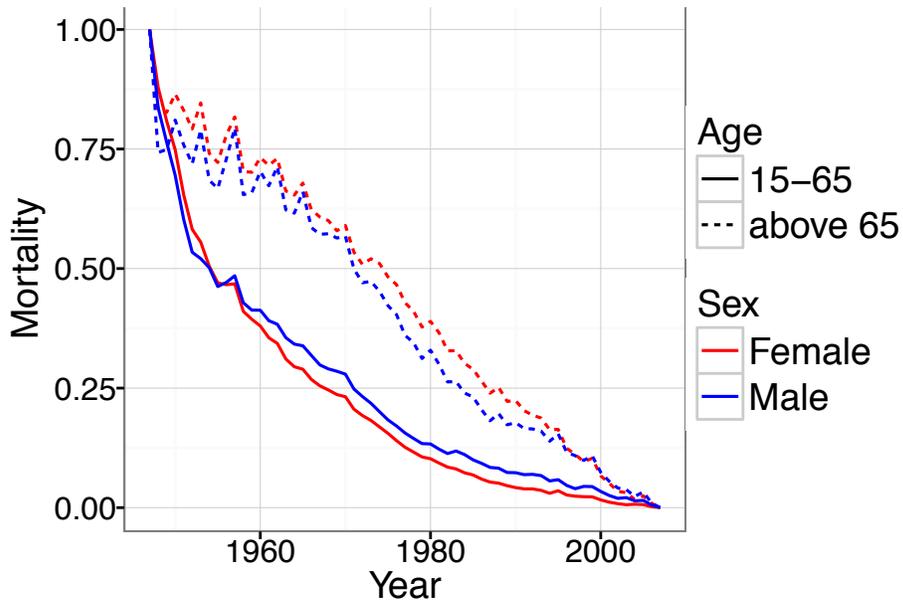

1   **Table S1** Regional data series in the Human Mortality Database used in our international

2   comparisons. The earliest and latest years show the data availability for each series in the

3   period 1947–2011

| Country or region | Earliest year | Latest year |
|---|---|---|
| Australia | 1947 | 2009 |
| Austria | 1947 | 2010 |
| Belgium | 1947 | 2009 |
| Bulgaria | 1970 | 2010 |
| Belarus | 1970 | 2010 |
| Canada | 1947 | 2007 |
| Switzerland | 1947 | 2011 |
| Chile | 1992 | 2005 |
| Czech Republic | 1950 | 2011 |
| West Germany | 1956 | 2010 |
| East Germany | 1956 | 2010 |
| Denmark | 1947 | 2011 |
| Spain | 1947 | 2009 |
| Estonia | 1959 | 2010 |
| Finland | 1947 | 2009 |
| France | 1947 | 2010 |
| England & Wales | 1947 | 2009 |
| Northern Ireland | 1947 | 2009 |
| Scotland | 1947 | 2009 |
| Hungary | 1950 | 2009 |
| Ireland | 1950 | 2009 |
| Iceland | 1947 | 2010 |
| Israel | 1983 | 2009 |
| Italy | 1947 | 2009 |
| Japan | 1947 | 2009 |
| Latvia | 1970 | 2010 |
| Luxembourg | 1960 | 2009 |
| Lithuania | 1959 | 2010 |
| Netherlands | 1947 | 2009 |
| Norway | 1947 | 2009 |
| New Zealand non-Maori | 1947 | 2008 |
| Poland | 1958 | 2009 |
| Portugal | 1947 | 2009 |
| Russia | 1959 | 2010 |
| Slovakia | 1950 | 2009 |
| Slovenia | 1983 | 2009 |
| Sweden | 1947 | 2011 |
| Taiwan | 1970 | 2010 |
| Ukraine | 1970 | 2009 |
| USA | 1947 | 2010 |